\documentclass[12pt, a4paper]{article}

\usepackage[pdftex]{graphicx}
\usepackage{bm,afterpage,amsmath,placeins,flafter,authblk,adjustbox,lscape,caption}
\usepackage[english]{babel}
\usepackage{setspace}
\usepackage{booktabs}
\usepackage{changes}
\usepackage[top=1.3in,left=1.1in,right=1.1in,bottom=1.3in]{geometry}
\usepackage{color}
\usepackage{hyperref}
\usepackage{epstopdf}
\graphicspath{{figures/}}

\date{}

\begin{document}

	\title{Players Movements and Team Shooting Performance: a Data Mining approach for Basketball.}
	\author{Rodolfo Metulini}
	\affil[1]{\small Department of Economics and Management, University of Brescia. Contrada Santa Chiara, 50, 25122 Brescia BS, Italy. }
	\maketitle

\begin{abstract}
In the domain of Sport Analytics, Global Positioning Systems devices are intensively used as they permit to retrieve players' movements. Team sports' managers and coaches are interested on the relation between players' patterns of movements and team performance, in order to better manage their team. In this paper we propose a Cluster Analysis and Multidimensional Scaling approach to find and describe separate patterns of players movements. Using real data of multiple professional basketball teams, we find, consistently over different case studies, that in the defensive clusters players are close one to another while the transition cluster are characterized by a large space among them. Moreover, we find the pattern of players' positioning that produce the best shooting performance.
\end{abstract}

\vskip 3mm

\vline

\textit{\textbf{Please cite as}: Metulini, R. (2018), Players Movements and Team Shooting Performance: a Data Mining approach for Basketball, ”49th Scientific meeting of the Italian Statistical Society" SIS2018 proceeding.}

\section{Introduction}
\label{sec:intro}

Studying the interaction between players in the court, in relation to team performance, is one of the most important issue in Sport Science, as team sports' Managers, more and more in recent years, are becoming aware of the potential of Data Analytics in order to better manage their team.  Recent years make it possible, thanks to the advent of Information Technology Systems (ITS), that permits to collect, store, manipulate and process a large amount of data. On the one hand, a sequence of relevant events of the match, such as passes, shots and fouls (player-specific) and time-outs (team-specific) takes the name of play-by-play. On the other hand, information on the movement of players on the court has been captured with the use of appropriate Geographical Positioning Systems (GPS) devices, for example the accelerometer, a device that measures proper velocity and positioning. Analysing players' interaction, however, is a complex task, as the trajectory of a single player depends on a large amount of factors related, among others, to coaches, single players and the whole team. The trajectory of a player depends on the trajectories of all the other players in the court, both teammates and opponents. Players interactions have been mainly studied in the new domain of ecological dynamics \cite{travassos2013performance,passos2016performance}. Typically, there are certain role definitions in a sports team that influence movements. Predefined strategies are used by the coach to achieve specific objectives. A common method to approach with this complexity in team sport analysis consists on segmenting a match into phases, as it facilitates the retrieval of significant moments of the game. For example  Perin et al. \cite{perin2013soccerstories} developed a system for visual exploration of phases in football, while, to the same goal, Metulini \cite{metulini2017spatio} propose motion charts. Cluster analysis methodology is widely used in team sports literature. To name a few, Sampaio and Janeira \cite{Sampaio2003statistical} applied a cluster analysis to investigate the discriminatory power of game statistics between winning and losing teams in the Portuguese Professional Basketball League, by using game final score differences, Ross \cite{ross2007segmenting} uses cluster analysis to segment team sport spectators identifying potential similarities according to demographic variables. Csataljay et al. \cite{Csataljay2009performance} used cluster approach to the purpose of identifying those critical performance indicators that most distinguish between winning and losing performances. However, differently from the aforementioned papers, to the aim of segmenting game into phases, in this paper we cluster time instants. In doing so, we use GPS tracked data. In this regard, Gonccalves \cite{gonccalves2018collective} applied a two-step cluster to classify the regularity in teammates dyads’ positioning. 
Metulini et al. \cite{metulini2017space} used cluster analysis to an amatorial basketball game in order to split the match in a number of separate time-periods, each identifying homogeneous spatial relations among players in the court. They also adopt a Multidimensional Scaling to visually characterize clusters and analysed the switch from \textit{defense} to \textit{offense} clusters, by mean of transition probabilities. This paper aims to fill the gap in Metulini et al., by extending the analysis to multiple matches. Moreover: i) we apply our cluster analysis procedure to professional basketball games, ii) we use the data generated by the algorithm proposed in Metulini \cite{metulini2017filtering} in order to consider active game moments only, iii) we use a more detailed labelling scheme introducing \textit{transition} moments, which permits a better interpretation of the transition probabilities. Last, we  characterize clusters in term of team performance, by retrieving shooting events throughout a video analysis.   

\section{Data and Methods}
\label{sec:meth}

Basketball is a sport generally played by two teams of five players each on a rectangular court. The objective is to shoot a ball through a hoop 46 centimeters in diameter and mounted at a height of 3.05 meters to backboards at each end of the court. According to FIBA rules, the match lasts 40 minutes, divided into four periods of 10 minutes each. There is a 2-minutes break after the first quarter and after the third quarter of the match. After the first half, there is a 10 to 20 minutes half-time break. In this paper we use tracked data from three games played by Italian professional basketball teams, at the Italian Basketball Cup Final Eight. MYagonism  (\url{https://www.myagonism.com/}) was in charge to set up a system to capture these data during the games, trough accelerometer devices. Each player worn a microchip that, having been connected with machines built around the court, collected the player's position (in pixels of 1 $cm^2$ size) in the $x$-axis (court length), the $y$-axis (court width), and in the $z$-axis (height). Data, filtered with a Kalman approach, has been detected at a millisecond level. Available data contain information on players' positioning, velocity and acceleration during the full game length. Throughout the text we will call the three games case study 1 (CS1), case study 2 (CS2) and case study 3 (CS3). As the initial dataset is provided to us considering the full game length, we cleaned it by dropping the pre-match, the quarter- and the half-time intervals and the post match periods, as well as the time-outs and the moments when a player is shooting a free-throw. More information on this filtering procedure can be found in Metulini \cite{metulini2017filtering}. The final dataset for CS1 counts for $206,332$ total rows, each identifying the milliseconds in which the system captured at least one player. CS2 dataset counts for $232,544$ rows, while CS3 counts for a total of $201,651$ rows.

We apply a $k$-means Cluster Analysis in order to group a set of objects. Cluster analysis is a method of grouping a set of objects in such a way the objects in the same group (clusters) are more similar to each other than to those in other groups. In our case, the objects are represented by the time instants, expressed in milliseconds, while the similarity is expressed in terms of distance between players' dyads. In the analyses that follows we only consider moments when a particular lineup is on the court. More specifically, we only consider lineups that played for at least 5 minutes. According to this criteria, we consider two lineups (\textit{p1, p3, p6, p7, p8} and \textit{p1, p4, p5, p7, p10}) for CS1, two (\textit{p1, p2, p4, p5, p6} and \textit{p1, p2, p5, p6, p8}) for CS2, and one lineup for CS3 (\textit{p2, p5, p6, p9, p10}, \textit{p} stays for player). We chose number of clusters based on the value of the between deviance (BD) / total deviance (TD) ratio and the increments of this value by increasing the number of clusters by one. We consistently, and surprisingly, find $k$=6 (BD/TD= around 45\%  along the different lineups, and relatively low increments for increasing $k$, for $k$$\ge$6) for almost all the lineups considered. Specifically, increasing the number of clusters from 5 to 6, BD/TD increments by around 11-12 \% in all the five lineups, while increasing from 6 to 7, BD/TD increments by around 6-7 \%.   

\section{Results}
\label{sec:res}
In this section we describe clusters for their dimension and their characteristics in term of pattern of player's positioning and team performance, along the five lineups.
According to the first lineup of CS1, the first cluster (C1) embeds 13.31\% of the observations (i.e. 13.31\% of the total game time), the other clusters, named C2, ..., C6, have size of 19.76\%, 3.40\%, 29.80\%, 6.41\% and 27.31\% of the total sample size, respectively. Consistently for all the five lineups, we find a couple of small clusters, with less than 10\% of the total observations, and 2-3 larger ones, containing at least 20\% of the observations. 

\begin{figure}[!htbp]
	\includegraphics[scale=0.6]{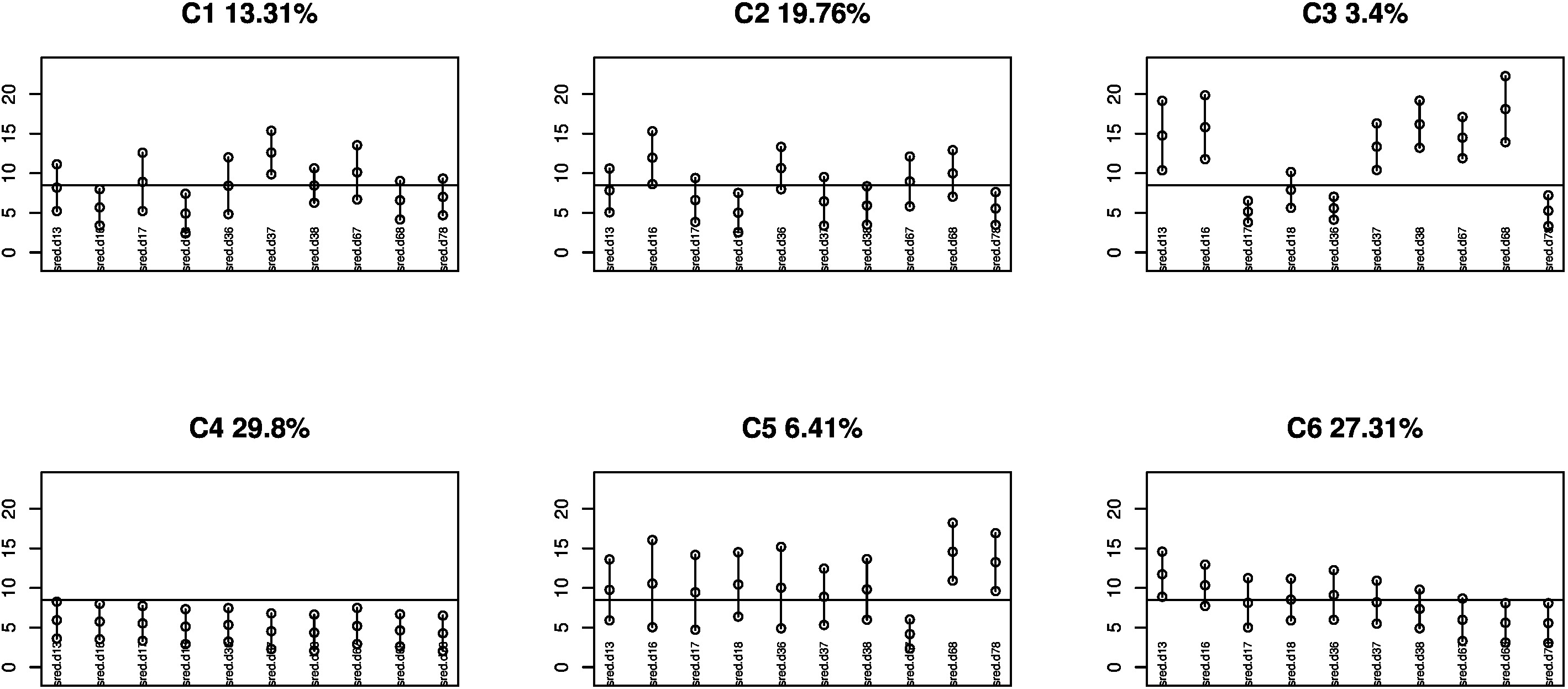}
	\caption{Profile plots representing, for each of the 6 clusters, the average distance among players' dyads.}
	\label{fig:profplot} 
\end{figure}

Cluster profile plots have been used to better interpret the players' spacing structure in each group. Figure \ref{fig:profplot} reports profile plot for the first lineup of CS1, to characterize groups in terms of average distances among players. In this case, we find the smaller cluster (C3, 3.4\% of observations) displaying large average distances among players (horizontal lines in Figure \ref{fig:profplot} represent the average value along the game time played by that  lineup). On the contrary, the larger cluster (C4, 29.8\%) displays all the average distances below the game average.  These two facts are confirmed in the second lineup of CS1, as it presents the larger cluster (C5, 40.4\%, which is not reported for the sake of space saving) displaying really small average distances, while its smaller cluster (C6, 3.2\%) reports large average distances. Same evidences have been found in other case studies. 

\begin{figure}[!htbp]
	\includegraphics[scale=0.6]{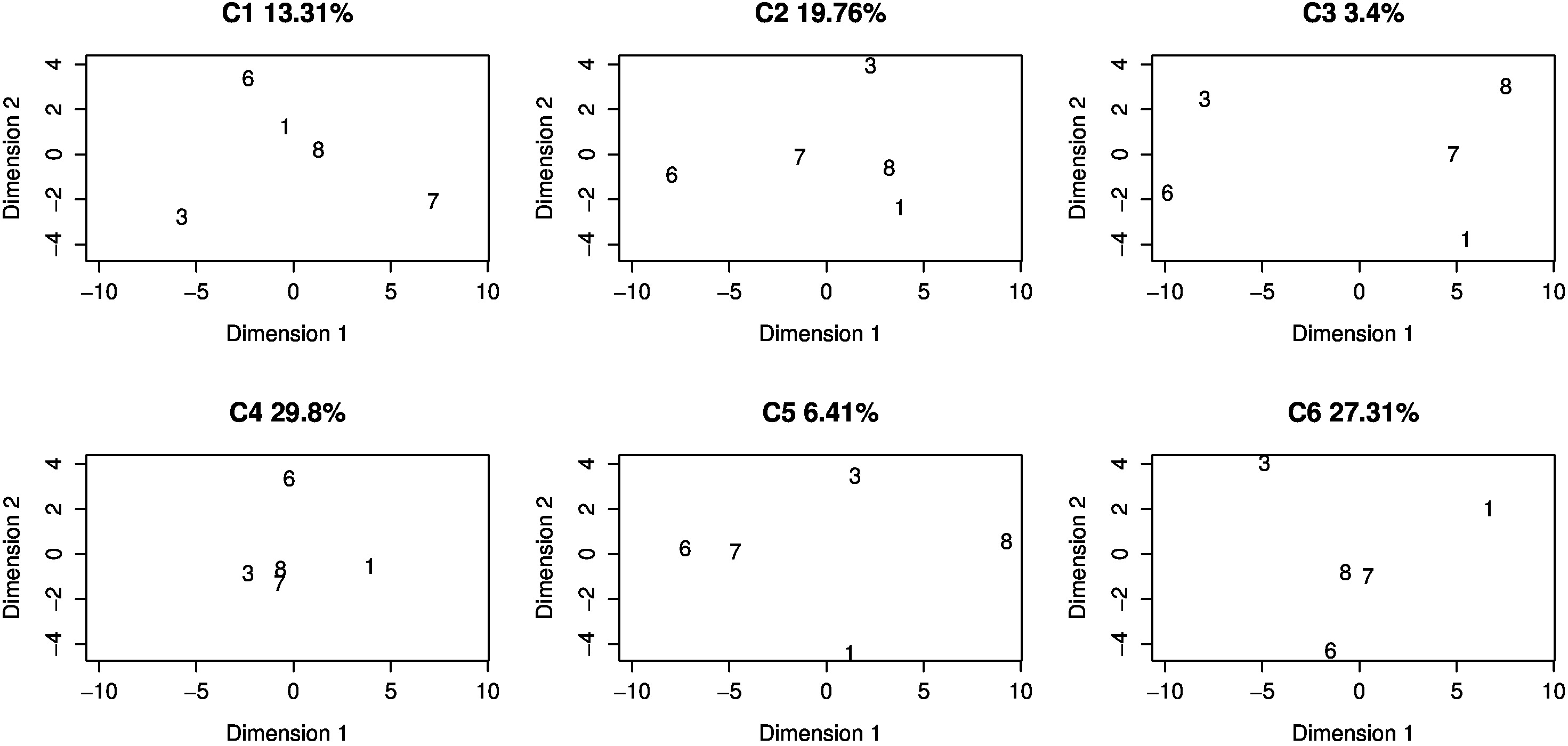}
	\caption{Map representing, for each of the 6 clusters, the average position in the $x-y$ axes of the five players in the court, using MDS.}
	\label{fig:mds}
\end{figure}

To the aim of producing further visual evidences, we used Multidimensional Scaling (MDS), which plots the differences between the groups in terms of positioning in the court. With MDS algorithm we aim to place each player in $N$-dimensional space such that the between-player average distances are preserved as well as possible. Each player is then assigned coordinates in each of the $N$ dimensions. We choose $N$=2 and we draw the related scatterplots. Figure \ref{fig:mds} reports the scatterplot for the first lineup of CS1. We observe strong differences between the positioning pattern among groups. The figure highlights large space among players in CS3, as also highlighted by the average distances in the profile plot. Moreover, moments in C4 are characterized by close to each others players. Despite not reported here, other lineups display similar MDS results: smaller clusters are characterized by large average distances and by a large space among players, while larger clusters by small average distances and by close to each others players.

The filtered datasets label each moment as \textit{offense}, \textit{defense} or \textit{transition} by mean of looking to the average $x$-axis positioning of the five players on the court. A moment is labelled as \textit{transition} when the average $x$-axis is in within the interval [-4,+4], where 0 corresponds to the half court line. Throughout this information, we associate each cluster to offense, defense or to transition, according how many time instants in a specific cluster corresponds to the specific label. 

\begin{table}[h!]
	\centering
	\begin{tabular}{c|ccccccc}
\textbf{Cluster}& &	\textbf{1}	&\textbf{2}&\textbf{	3}&	\textbf{4}&	\textbf{5}&	\textbf{6}\\
\hline
&&&&&&&\\
\textbf{TR}& &	8.41&	21.76	&\textbf{82.11}&	7.08&	\textbf{54.49}&	10.53\\
\textbf{D}&&	22.74&	10.28&	6.6&	\textbf{70.48}	&23.98&	17.95\\
\textbf{O}&	&\textbf{68.85}&	\textbf{67.97}&	11.29&	22.45&	21.53&	\textbf{71.52}\\
\hline
&&&&&&&\\
\textbf{Total} &  &100.00 &100.00 &100.00 &100.00 &100.00 &100.00 \\
	\end{tabular}	
	\caption{Percentages of time instants classified in Transition (TR) Defense (D) or Offense (O), for each cluster.}
	\label{tab:lab}
\end{table}

Table \ref{tab:lab} reports related percentages for the first lineup in CS1. Clusters C1, C2 and C6 mainly correspond to offense (respectively, for the 68.85\%, 67.97\% and 71.52\% of the times), C3 and C5 correspond to defensive actions (82.11\% and 54.49\% of the times, respectively), while C4 corresponds to defense (70.48\%). It emerges that large clusters with small average distances among players contains defensive moments. Moreover, the small cluster with large distances corresponds to transition. This result is consistent in all the five considered lineups. For example, in the second lineup of CS1, the small cluster (C6) corresponds to transition moments for the 80.76\% of the time. The large cluster with corresponding small distances (C5) contains moments classified as defense the 72.99\% of the times.

Table \ref{fig:transmat_ad} shows the transition matrix for the first lineup of CS1, which reports the relative frequency in which subsequent moments in time report a switch from a cluster to a different one. 
Main diagonal values of the matrix have been set to zero, so that each column percentages sum to 100\% without including subsequent moments in which there isn't a switch. 

\begin{table}[h!]
	\centering
	\begin{tabular}{c|ccccccc}
		\textbf{Cluster label} &   & \textbf{ C1} & \textbf{C2} & \textbf{C3} & \textbf{C4}  & \textbf{C5} & \textbf{C6} \\
		\hline
		&&&&&&&\\
		\textbf{C1} & & 0.00 &11.27&   10 & 8.45 &  15 &10.34\\
		\textbf{C2} &&31.03 & 0.00 &  10& 23.94  & 15 &35.34\\
		\textbf{C3}&& 0.00 & 1.41  &  0 & 0.00&    0  &7.76\\
		\textbf{C4} &&34.48 &21.13   & 0 & 0.00&   25 &35.34\\
		\textbf{C5}& & 3.45  &4.23  &  0 & 4.23&   0 &11.21\\
		\textbf{C6} &&31.03 &61.97 &  80& 63.38 &  45 & 0.00\\
	\end{tabular}	
	\caption{Transition matrix reporting the relative frequency subsequent moments  ($t$, $t + 1$) report a switch from a group to a different one.}
	\label{fig:transmat_ad}
\end{table}

It emerges that, for the 34.48\% of the times C1 switches to a new cluster, it switches to C4. It also switch to C2 and C6, respectively for the 31.03\% and for the 31.03\% of the times. We can note that C2, C4 and C6 are the three largest clusters. C3, marked as \textit{Transition}, switches 80\% of the times to C6 (a offensive cluster). Moreover, C2 switches most of the times (61.97\%) to C6. C2 and C6 are both marked as offensive clusters. Since the total number of switches for this lineup is equal to 309, and this lineup played for a total of 8 minutes and 21 seconds, on average we have a switch every 2 seconds. For this reason we have frequent cluster switches during the same action. Switch from C2 to C6 is an example: players change patterns of positioning during the same action. Table \ref{sec:conc} highlights that offensive clusters often switch to another offensive cluster (beside C2, C1 switches for the 31.03\% of the times to C2 and for the 31.03\% of the times to C6, C6 switches for the 35.34\% of the times to C2). This evidence is confirmed in the other case studies, since, in the second lineup of CS1, we have three offensive clusters (C1, C2 and C3); C1 switches to C2 for the 33.33\% of the times, C3 switches to C2 for the 40.91\% of the times. In the first lineup of C2, we find three offensive clusters (C3, C4 and C6); C4 switches to C6 for the 75.86\% of the times. 

\vline

With this in mind, we can not associate a cluster with a whole action played with a particular tactic, instead, we have to interpret offensive clusters as subsequent players' positioning configurations, to the aim of finding the best positioning for a good shot.

In light of this, we collect the shooting events of the match. Since play-by-play data are not available for this tournament, we collect such events by watching the video of the game. Zuccolotto et al. \cite{zuccolotto2017big} analysed the shooting performance under pressure. Here we study shooting performance with respect to different players' positioning patterns, by associating shots to the cluster in which the team was (at the moment of the shot). We take into consideration only shots from the court, disregarding free throws. During the 8 minutes and 21 seconds players \textit{p1, p3, p6, p7} and \textit{p8} were in the court together, the team made 15 shots from the court, with 7 made shots and 8 missed, for a percentage of 46.67\%. 
We find that most of these shots (8) has been attempted in moments that belongs to cluster C6. During this cluster, the team scored 5 out of 8 total attempts, with a really high percentage of 62.5\%, ways higher than the average of 46.67\%.  Moreover, the team attempted only 4 shots (2 of them made) during cluster C1, only 2 shots (both missed) during cluster C2 and it missed a shot during cluster C5. So, 14 out of 15 shots have been attempted during the clusters labelled as offensive (i.e. C1, C2 and C6) while only one during a transition cluster (C5). Looking to bottom-right chart in Figure \ref{fig:mds} (C6), we find player 3 far away from the others. We could suppose that the tactic of the team was to leave that player free to shot on the weaker side of the court. Results support the idea that C6 represents the cluster of (good) shooting moments. Furthermore, the other offensive (C1, C2) and transition (C3, C5) clusters often switch to cluster C6, which support our hypothesis of subsequent game configurations to the aim of finding the best positioning for a good shot: the best positioning to shot is that in C6 moments. 

\section{Conclusions}
\label{sec:conc}
In recent years, the availability of `big data" in Sport Science increased the possibility to extract insights from the games that are useful for managers and coaches, as they are interested to improve their team's performances. In particular, with the advent of Information Technology Systems, the availability of players' trajectories permits to analyse the space-time patterns with a variety of approaches. With this paper we pursue the points raised by Metulini et al. \cite{metulini2017space} as suggestions for future research, by analyzing multiple professional games and relate clusters with team shooting performance. We segmented the game into phases of play and we characterized each phase in terms of spacing structure among players, relative distances among them and whether they represent an offensive, a defensive or a transition play, finding substantial differences among different phases. Moreover, we analysed this structure in terms of shooting performance, finding the cluster corresponding to the best shooting performance. These results shed light on the potentiality of data-mining methods for players' movement analysis in team sports. In future research we aim to better explain the relation between players' positioning and team performance, adding more play-by-play data and analysing this relationship for a larger amount of time and for multiple matches.

\section{Acknowledgement}
Research carried out in collaboration with the Big\&Open Data Innovation Laboratory (BODaI-Lab), University of Brescia (project nr. 03-2016, title Big Data Analytics in Sports, \url{www.bodai.unibs.it/BDSports/}), granted by Fondazione Cariplo and Regione Lombardia. Authors would like to thank MYagonism (\url{https://www.myagonism.com/}) for having provided the data and Paola Zuccolotto (University of Brescia) for the useful discussions.

%
%

\begin{thebibliography}{99.}%
%
%


\bibitem{travassos2013performance} Travassos, B., Davids, K., Araujo, D., Esteves, P. T.: Performance analysis in team sports: Advances from an Ecological Dynamics approach. International Journal of Performance Analysis in Sport 13.1 (2013): 83-95.
%
\bibitem{passos2016performance} Passos, P., Araujo, D., Volossovitch, A.: Performance Analysis in Team Sports. Routledge (2016).


\bibitem{perin2013soccerstories} Perin, C., Vuillemot, R., Fekete, J. D.: SoccerStories: A kick-off for visual soccer analysis. IEEE transactions on visualization and computer graphics 19.12 (2013): 2506-2515.
%

\bibitem{metulini2017spatio} Metulini, R.: Spatio-Temporal Movements in Team Sports: A Visualization approach using Motion Charts. Electronic Journal of Applied Statistical Analysis Vol. 10.3 (2017): 809-831.

\bibitem{Sampaio2003statistical} Sampaio, J., Janeira, M.: Statistical analyses of basketball team performance: understanding teams’ wins and losses according to a different index of ball possessions. International Journal of Performance Analysis in Sport 3.1 (2003): 40-49.

\bibitem{ross2007segmenting} Ross, S. D.: Segmenting sport fans using brand associations: A cluster analysis. Sport Marketing Quarterly, 16.1 (2007): 15.

\bibitem{Csataljay2009performance} Csataljay, G., O’Donoghue, P., Hughes, M., Dancs, H.: Performance indicators that distinguish winning and losing teams in basketball. International Journal of Performance Analysis in Sport 9.1 (2009): 60-66.


\bibitem{gonccalves2018collective} Gonçalves, B. S. V.: Collective movement behaviour in association football. UTAD Universidade de Tras-os-Montes e Alto Douro (2018)


\bibitem{metulini2017space} Metulini, R., Marisera, M., Zuccolotto, P.: Space-Time Analysis of Movements in Basketball using Sensor Data. Statistics and Data Science: new challenges, new generations SIS2017 proceeding. Firenze Uiversity Press. eISBN:
978-88-6453-521-0 (2017).

\bibitem{metulini2017filtering} Metulini, R.: Filtering procedures for sensor data in basketball. Statistics\&Applications 2 (2017).

\bibitem{zuccolotto2017big} Zuccolotto, P., Manisera, M., Sandri, M.: Big data analytics for modeling scoring probability in basketball: The effect of shooting under high-pressure conditions. International Journal of Sports Science \& Coaching (2017): 1747954117737492.





\end{thebibliography}
%

\end{document}